\documentclass[preprint2]{aastex}

\slugcomment{}

\shorttitle{Infrared study of very fast nova V2468 Cyg}
\shortauthors{Raj et al.}

\begin{document}

%% LaTeX will automatically break titles if they run longer than
%% one line. However, you may use \\ to force a line break if
%% you desire.

\title{IR study of nova V2468 Cyg from early decline \\
    to the coronal phase}

%% Use \author, \affil, and the \and command to format
%% author and affiliation information.
%% Note that \email has replaced the old \authoremail command
%% from AASTeX v4.0. You can use \email to mark an email address
%% anywhere in the paper, not just in the front matter.
%% As in the title, use \\ to force line breaks.

\author{A. Raj\altaffilmark{}, 
\affil{Korea Astronomy and Space Science Institute, Daejeon, 305-348, Korea.}
N. M. Ashok\altaffilmark{}, 
\affil{Physical Research Laboratory, Navarangpura, Ahmedabad, India.}
Richard J. Rudy\altaffilmark{}\footnote{Visiting Astronomer at the Infrared Telescope Facility, which is operated by the University of Hawaii under Cooperative 
Agreement no. NNX-08AE38A with the National Aeronautics and Space Administration, Science Mission Directorate, Planetary Astronomy Program.},
Ray W. Russell\altaffilmark{1},  
David K. Lynch\altaffilmark{1}, 
Charles E. Woodward\altaffilmark{1}, 
Michael Sitko\altaffilmark{1}, 
Amanda Day-Wilson\altaffilmark{1}, 
 \\
\affil{IRTF, NASA, USA}
R. Brad Perry\altaffilmark{},
\affil{}
Sang Chul KIM\altaffilmark{}
and Mina Pak\altaffilmark{}
\affil{Korea Astronomy and Space Science Institute, Daejeon, and Korea University of Science and Technology (UST), 305-348, Korea.}
}

% \author{C. D. Biemesderfer\altaffilmark{4,5}}
% \affil{National Optical Astronomy Observatories, Tucson, AZ 85719}
\email{ashish@kasi.re.kr}

% \and

% \author{R. J. Hanisch\altaffilmark{5}}
% \affil{Space Telescope Science Institute, Baltimore, MD 21218}

%% Notice that each of these authors has alternate affiliations, which
%% are identified by the \altaffilmark after each name.  Specify alternate
%% affiliation information with \altaffiltext, with one command per each
%% affiliation.
% 
% \altaffiltext{1}{Visiting Astronomer, Cerro Tololo Inter-American Observatory.
% CTIO is operated by AURA, Inc.\ under contract to the National Science
% Foundation.}
% \altaffiltext{2}{Society of Fellows, Harvard University.}
% \altaffiltext{3}{present address: Center for Astrophysics,
%     60 Garden Street, Cambridge, MA 02138}
% \altaffiltext{4}{Visiting Programmer, Space Telescope Science Institute}
% \altaffiltext{5}{Patron, Alonso's Bar and Grill}

%% Mark off your abstract in the ``abstract'' environment. In the manuscript
%% style, abstract will output a Received/Accepted line after the
%% title and affiliation information. No date will appear since the author
%% does not have this information. The dates will be filled in by the
%% editorial office after submission.

\begin{abstract}
We present infrared spectroscopic and photometric observations of the nova V2468 Cyg covering the period from 2008 March 13 till 2008 November 11. 
The $JHK$ spectra of the object have been taken from the Mount Abu Infrared Observatory using the Near-Infrared Imager/Spectrometer.  Spectra from 0.8-5.2 $\mu$m are 
also presented that were obtained using the NASA Infrared Telescope Facility and the SPEX instrument. The spectra are dominated by strong
 H {\sc i} lines from the Brackett and Paschen series, Fe\,{\sc ii}, O\,{\sc i} and C\,{\sc i} lines in the initial days, typical of an Fe II type nova.
The lines were broader in the period immediately after outburst with measured FWHM of 1800-2300 km s$^{-1}$ for the Pa$\beta$ and Br$\gamma$ 
lines. These values narrowed to 
1500-1600 km s$^{-1}$ by 12 days from outburst. The spectra showed prominent He {\sc i} lines at 1.0830 and 
2.0581 $\mu$m together with H {\sc i}
and O {\sc i} emission features after 36 days from outburst. Our IR observations show the comparatively broad emission lines, the rapid development of the spectrum 
to higher ionization, the early appearance of coronal lines, and the absence of dust emission, all features that indicate the hybrid nature of the nova. 
This is perhaps the most extensively observed example of a probable Fe IIb type nova at the infrared wavelengths. We also notice a short lived emission line of Fe II at 2.0888$\mu$m which was present between April 9, 2008 to May 9, 2008.  
No dust emission is seen from the nova ejecta. We have also estimated the range for the ejecta mass in V2468 Cyg to be 3 $\times$ 10$^{-6}$ - 10$^{-5}$M$_\odot$. 
\end{abstract}

%% Keywords should appear after the \end{abstract} command. The uncommented
%% example has been keyed in ApJ style. See the instructions to authors
%% for the journal to which you are submitting your paper to determine
%% what keyword punctuation is appropriate.


%% From the front matter, we move on to the body of the paper.
%% In the first two sections, notice the use of the natbib \citep
%% and \citet commands to identify citations.  The citations are
%% tied to the reference list via symbolic KEYs. The KEY corresponds
%% to the KEY in the \bibitem in the reference list below. We have
%% chosen the first three characters of the first author's name plus
%% the last two numeral of the year of publication as our KEY for
%% each reference.

%% Authors who wish to have the most important objects in their paper
%% linked in the electronic edition to a data center may do so by tagging
%% their objects with \objectname{} or \object{}.  Each macro takes the
%% object name as its required argument. The optional, square-bracket 
%% argument should be used in cases where the data center identification
%% differs from what is to be printed in the paper.  The text appearing 
%% in curly braces is what will appear in print in the published paper. 
%% If the object name is recognized by the data centers, it will be linked
%% in the electronic edition to the object data available at the data centers  
%%
%% Note that for sources with brackets in their names, e.g. [WEG2004] 14h-090,
%% the brackets must be escaped with backslashes when used in the first
%% square-bracket argument, for instance, \object[\[WEG2004\] 14h-090]{90}).
%%  Otherwise, LaTeX will issue an error. 

\section{Introduction}

Nova V2468 Cyg was discovered on 2008 March 7.801 UT by Hiroshi Kaneda on nine 4s unfiltered CCD frames with limiting magnitude 
10.7 at $V$ = 8.2 $\pm$ 0.3 (Nakano \& Kaneda 2008). Nothing was visible at the same position on his three patrol images taken on 2007 October 5, 2008 January 1 and 2008 February 18 with limiting magnitude 10.7.
A low resolution optical spectrum obtained on March 8.794 UT with a 1.88-m telescope (+ KOOLS) at the Okayama Astrophysical Observatory by Nogami, Kuriyama \& Iwata
(2008) showed a blue continuum with strong Balmer and Fe II lines having prominent P Cygni profiles. The FWHM of the emission component of H$\alpha$ was about
1000 km s$^{-1}$, and the absorption component was blue shifted from the emission peak by about 880 km s$^{-1}$. They suggested that the object was a classical nova before or 
around optical maximum.
The next spectrum taken by Beaky (2008) on March 11.46 UT, showed prominent emission lines of H$\alpha$ and H$\beta$ and large Fe II emission lines which confirmed
that
the nova was of the Fe II class. The near-IR observations reported by Ashok \& Banerjee (2008) that were taken on March 14, also showed that the spectra are typical
of a classical nova showing prominent H I emission lines of Pa$\beta$, Pa$\gamma$ and Br$\gamma$ and other
Brackett-series lines. The other prominent features seen were O I lines at 1.1287 and 1.3164 $\mu$m, moderately strong C I lines in the J
band, and the Na I 2.2056 and 2.2084 $\mu$m lines. The FWHM of the infrared H I lines were in the range 2000-2200 km s$^{-1}$. Rudy et al. (2008) reported on the IR spectra 
for the nova on two epochs: 2008 March 13 and 2008 April 12 UT. Among other things, they found that the Ca II infrared triplet, which was strong in March, was almost undetectable 
in the April spectrum. X-ray emission from V2468 Cyg was also detected in 2009 (Schwarz et al. 2009), 2011 (Schwarz et al. 2011) and 
2012 (Page et al. 2012). 

\section{Observations}

Near-IR observations were obtained using the 1.2m telescope of Mt. Abu Infrared Observatory and NASA Infrared Telescope Facility from 2008 March 13 to 2008 
November 11. The log of the spectroscopic and photometric observations is given in Table 1. The spectra obtained at Mt. Abu at a resolution of $\sim$ 1000 were 
using the Near-Infrared
Imager/Spectrometer incorporating a 256 $\times$ 256 HgCdTe NICMOS3 array. The IRTF data had a resolution from 1000-2000 and were acquired with the SPEX instrument, 
a cross dispersed imaging spectrometer incorporating a 1024 $\times$ 1024 InSb detector and covering the wavelength range 0.8-5.5 $\mu$m (Rayner et. al. 2003). 
In each of the $JHK$ bands a set of spectra was taken with the nova 
off-set to
two different positions along the slit which were subtracted from each other to remove sky and detector dark current contributions. Spectral calibration 
was done using the OH sky lines that register with the stellar spectra. The spectra of
the comparison star SAO 88071 were taken at similar airmass as that of V2468 Cyg to ensure that the ratioing process (nova spectrum divided by the
standard star spectrum) removes the telluric features reliably. To avoid artificially generated emission lines in the ratioed spectrum, the H I
absorption lines in the spectra of standard star were removed before ratioing. The ratioed spectra were then multiplied by a
blackbody curve corresponding to the standard star's effective temperature to yield the final spectra. Reduction of the IRTF data follows a similar process but uses early A stars and accounts for the profiles of  H I absorption lines.  
It is described by Cushing et. al. (2004) and Vacca et. a.l (2003). The A0V star HD 196724 was the standard star used for the observations of 2008 March 13, 
2008 August 21 and 2008 October 4 while the A0V star HD 192538 was the standard for 2008 April 12 and 2008 May 9.

\begin{table*}
\centering
\scriptsize
\caption{Log of the Mt. Abu and NASA Infrared Telescope Facility observations of V2468 Cyg. The $JHK$ band magnitudes are listed.}
\begin{tabular}{cccccccc}
\hline
Date of      &Spectroscopic Observations     & J& H & K& Observatory \& Instrument \\
Observation   &Bands (JHKLM)             &             &  & &  \\
%\hline
 %  &   &   &  \\
\hline
2008 Mar. 13    & JHK         &--    &--   &--&NASA IRTF SPEX\\
2008 Mar. 14    &JHK &--  &-- &--& Mt. Abu NICMOS3\\
2008 Mar. 15    &JHK    &6.21$\pm$0.06 &6.02$\pm$0.05 &5.55$\pm$0.05& Mt. Abu NICMOS3\\
2008 Mar. 19    &JHK   &-- &-- &-- &Mt. Abu NICMOS3\\
2008 Apr. 08    &-- &7.50$\pm$0.01 &7.50$\pm$0.02  &6.94$\pm$0.05&Mt. Abu NICMOS3\\
2008 Apr. 09    &JHK  &--  &-- &-- &Mt. Abu NICMOS3\\
2008 Apr. 12    &JHK  &--  &-- &--& NASA IRTF SPEX\\
2008 Apr. 22    &JHK &7.73$\pm$0.05 &7.92$\pm$0.03  &7.20$\pm$0.03& Mt. Abu NICMOS3\\
2008 May. 01    &-- &8.63$\pm$0.01 &8.75$\pm$0.02  &7.92$\pm$0.01& Mt. Abu NICMOS3\\
2008 May. 09    &JHKLM &--   &--   &--&NASA IRTF SPEX\\
2008 May. 15    &JHK  &--  &--&-- & Mt. Abu NICMOS3\\
2008 Aug. 21    &JHKLM  &--  &-- &-- &NASA IRTF SPEX\\
2008 Oct. 04    & JHKLM &--  &-- &-- &NASA IRTF SPEX\\
2008 Nov. 08    &-- &12.95$\pm$0.04 &12.77$\pm$0.11  &12.73$\pm$0.20&Mt. Abu NICMOS3\\
2008 Nov. 11    &-- &13.24$\pm$0.05 &13.27$\pm$0.18  &12.75$\pm$0.20&Mt. Abu NICMOS3\\

\hline
\end{tabular}
\label{table1}
\end{table*}

Photometry in the $JHK$ bands was done in clear sky conditions using the NICMOS3 array in the imaging mode. Several frames, in 4 dithered positions,
 offset by $\sim$ 30 arcsec were obtained in all the bands. The sky frames, which are subtracted from the nova frames, were generated by median
combining the dithered frames. The star SAO 68744 located close to the nova, was used for photometric calibration. Further details of the data 
reduction are described in Banerjee et al. (2014). The data were reduced and analyzed using $IRAF$.

\begin{table}
\scriptsize
\caption[]{A list of the lines identified from the $JHKLM$ spectra shown in Figures 2-4. All lines are seen at any one epoch. The
  additional lines contributing to the identified lines are listed.}
\begin{tabular}{llllll}
\hline\\
Wavelength & Species  & Other contributing  \\
(${\rm{\mu}}$m) & &lines and remarks   \\
\hline
\hline \\
0.8446   & O\,{\sc i}         &      \\
0.8863 & Pa 11 & \\
0.9013 & Pa 10  &  \\
0.9226 & Pa 9 & \\
0.9545 & Pa 8 &  \\
0.9911   & [Si \,{\sc viii}]     &           \\
1.0043 & Pa 7 & \\
1.0830   & He\,{\sc i} & \\
1.0938   & Pa $\gamma$        &      \\
1.1126   & u.i.                & Fe\,{\sc ii} ? \\
1.1287   & O\,{\sc i}         &      \\
% 1.1330   & C\,{\sc i}         &        \\
% 1.1381   & Na\,{\sc i}        &    C\,{\sc i} 1.1373  \\
% 1.1404   & Na\,{\sc i}        &    C\,{\sc i} 1.1415  \\
1.1626 & He\, {\sc ii}  & \\
1.1600-1.1674   & C\,{\sc i}  & the strongest lines at 1.1653,\\
                &             &           1.1659,1.16696    \\
1.1746-1.1800   & C\,{\sc i}  & the strongest lines at 1.1748,  \\
                &             &          1.1753,1.1755      \\
1.1828          & Mg\,{\sc i} &               \\
1.1819-1.2614   & several C\,{\sc i}  & the strongest lines at 1.1880,  \\
                & and N\,{\sc i}      &           1.1896 \\
1.1969 & He\,{\sc i}  & \\
1.2461,1.2469 & N\,{\sc i} & blended with O\,{\sc i} 1.2464 \\
1.2528 & He\,{\sc i}  & \\
1.2562,1.2569 & C\,{\sc i} & blended with O\,{\sc i} 1. 2570 \\
1.2818   & Pa $\beta$         &     \\
% 1.2950   & C\,{\sc i}         &     \\
1.3164   & O\,{\sc i}         &     \\
1.3450   & N\,{\sc i}         &     \\
1.4760 & He\,{\sc ii}  & \\
1.4882 & He\,{\sc i}  & \\
% 1.5040   & Mg\,{\sc i}        &  blended with Mg\,{\sc i} 1.5025,\\
%          &                    &         1.5048 \\
% 1.5256   & Br 19              &           \\
% 1.5341   & Br 18              &           \\
1.5439   & Br 17              &           \\
1.5557   & Br 16              &           \\
1.5701   & Br 15              &           \\
% 1.5749   & Mg\,{\sc i}        & blended with Mg\,{\sc i} 1.5741,  \\

%          &                    &           1.5766,C\,{\sc i} 1.5788 \\
1.5881   & Br 14              &  blended with C\,{\sc i} 1.5853    \\
% 1.6005   & C\,{\sc i}         &           \\
1.6109   & Br 13              &           \\
1.6407   & Br 12              &           \\
1.6806   & Br 11              &           \\
% 1.6890   & C\,{\sc i}         &           \\
1.7002   & He\,{\sc i}        &   \\
% 1.7045   & C\,{\sc i}         &           \\
1.7109   & Mg\,{\sc i}        &               \\
% 1.7234-1.7275 & C\,{\sc i}    & several C\,{\sc i} lines  \\
1.7362   & Br 10              &  affected by C\,{\sc i} 1.7339 line    \\
% 1.7449 & C\,{\sc i}           &           \\
% 1.7605-1.7638 & C\,{\sc i}    &           \\
% 1.7675  & u.i                 &           \\
1.7769-1.7814 & C\,{\sc i}    &           \\
% 1.8021  & O\,{\sc i} ?        &           \\
1.9445  & Br 8                &           \\
1.9645 & [Si \,{\sc vi}]     &           \\
1.9722  & C\,{\sc i}          &           \\
2.0449 & [Al\,{\sc ix}] & \\
2.0581 & He\,{\sc i}          &           \\
2.0888  & u.i                 & Fe\,{\sc ii} ?          \\
% 2.1023 & C\,{\sc i}           &           \\
2.1120, 2.1132  & He\,{\sc i}  &  \\
2.1156-2.1295 & C\,{\sc i}    &           \\
% 2.1452 & Na\,{\sc i}          &           \\
2.1655   & Br $\gamma$        &           \\
2.1882 & He\,{\sc ii}  & \\ 
2.2056,2.2084 & Na\,{\sc i}   &           \\
% 2.2156-2.2167 & C\,{\sc i}    &           \\
% 2.2520  & u.i     &    \\
2.2906 & C\,{\sc i}           &           \\
% 2.3130 & C\,{\sc i}           &           \\
2.3205 & [Ca \,{\sc viii}]     &           \\
2.4693  &              H\,{\sc i} 5-18 & \\
2.4946   &        H\,{\sc i} 5-17 &  \\
2.5254    &            H\,{\sc i} 5-16& \\
3.0384     &           H\,{\sc i} 5-10& \\
3.0908      &     He\,{\sc ii} 6-7& \\
3.2067       &         [Ca\,{\sc iv}]& \\
3.2961        &        H\,{\sc i} 5-9& \\
3.6060         &  H\,{\sc i} 6-20& \\
3.6449          & H\,{\sc i} 6-19& \\
3.6916           &H\,{\sc i} 6-18& \\
3.7395            &    H\,{\sc i} 5-8& \\
3.8184             &   H\,{\sc i} 6-16& \\
3.9065              &  H\,{\sc i} 6-15& \\
4.0512               & H\,{\sc i} 4-5& \\
4.6525                &H\,{\sc i} 5-7& \\
4.6712                &H\,{\sc i} 6-11& \\
5.1273                &H\,{\sc i} 6-10& \\

\hline
\end{tabular}
\end{table}

\section{Results}

\subsection{General characteristics of $JHK$ light curves}

\begin{figure*}
\begin{center}
\includegraphics[width=3.5in,height=4.0in,clip]{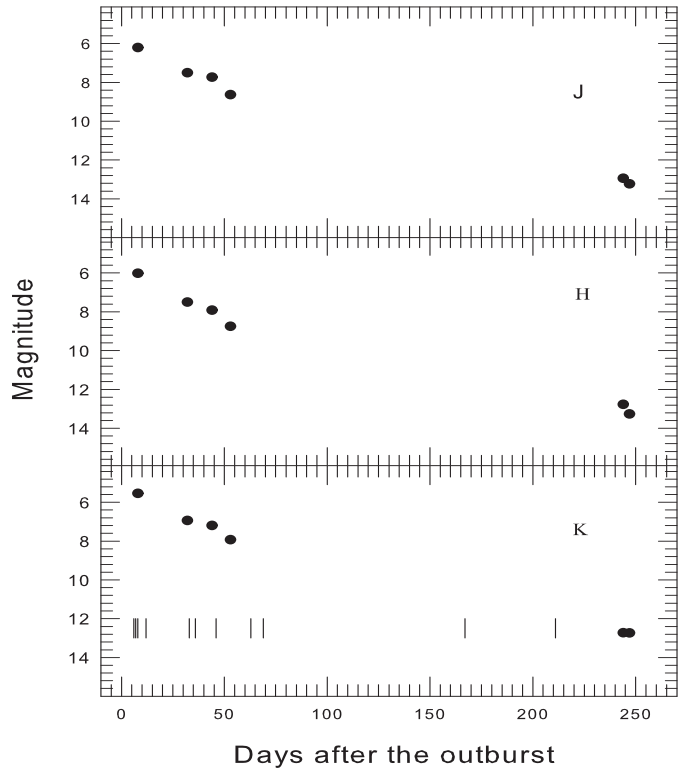}
\caption{The $JHK$ band light curves of V2468 Cyg, based on the data obtained from Mt. Abu, are presented. The spectroscopic observations are marked as solid lines in 
the lowest panel.}
%\label{ch5_2}
\end{center}
\end{figure*}

The light curves based on the $JHK$ magnitudes from Mt. Abu (see Table 1) 
are presented in Fig. 1. The spectroscopic observations for the nova are marked in the lowest panel of Fig. 1. 
The optical light curve shows small amplitude outbursts at 110, 180 and 240 days from the date of the outburst (see Fig. 1 of Tarasova 2013). There is no indication of dust formation in the nova ejecta at any 
stage of its evolution.  
The $JHK$ light curves show decline in the $JHK$ magnitudes and the observations taken in 2008 November clearly shows the cooling of 
the nova envelop.  

\subsection{Line identification, evolution and general characteristics of the $JHKLM$ spectra}

The $JHK$ spectra from Mt. Abu and NASA Infrared Telescope Facility are presented in Fig. 2 and 3, respectively. The $LM$ spectra from NASA Infrared Telescope Facility 
are presented in Fig. 4. The line identification is given in Table 2.
The infrared observations presented here cover the phase after optical maximum with the first infrared spectra taken on 2008 March 13.

The $JHK$ band spectra obtained on 2008 March 13 show the Ca II infrared triplet Pa$\gamma$, Fe II line at 1.1126 $\mu$m, strong line of OI at 1.1287 and 
1.3164 $\mu$m, 
CI line at 1.1748, 1.2569, 1.7808, 1.9722, 2.1186, 2.1220 and 2.2906 $\mu$m, weak line of Mg at 1.1828 and 1.7109 $\mu$m,
NI line at 1.2461 $\mu$m, Pa$\beta$ at 1.2818 $\mu$m, Brackett series 
lines and blend of Fe II and Br11 line, weak and broad 
He I line at 2.0581 $\mu$m, Na I line at 2.2056 $\mu$m. The FWHM value for the H I lines is between 1800-1900 km s$^{-1}$.
Similar lines have been seen in our previous observations of Fe II class of novae viz. V5579 Sgr (Raj et al. 2011), V496 Sct (Raj et al. 2012) and V5584 Sgr 
(Raj et al. 2014). 

The spectra taken on 2008 March 14 and 15 are similar to the previous spectra but we notice the change in the FWHM for the H I lines of about 
400 km s$^{-1}$ (see Fig. 6). This may be because of the very small amplitude outburst close to these epochs (Tarasova 2013).

\begin{figure*}
 \begin{center}
% \centering
 \includegraphics [width=5.5in,height=5.5in]{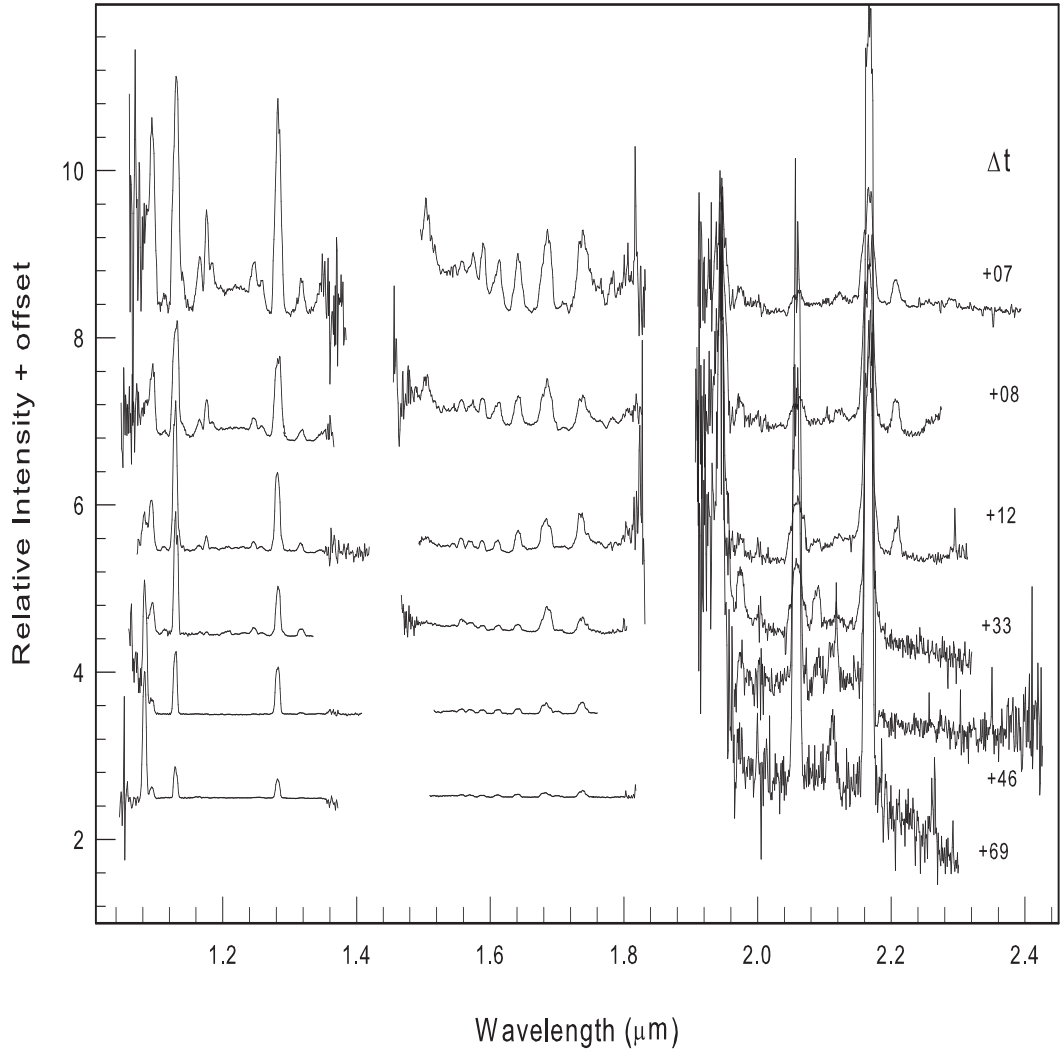}
  \caption{The $JHK$ band spectra of V2468 Cyg are shown at different epochs. These observations have been taken from Mt. Abu IR telescope. 
  The relative intensity is normalized to
unity at 1.25 ${\rm{\mu}}$m, 1.65 ${\rm{\mu}}$m and 2.2 ${\rm{\mu}}$m for $JHK$ bands, respectively. The time from outburst is given for each spectrum.}
 %\label{ch5_3}
 \end{center}
 \end{figure*}
 
The spectra taken on 2008 March 19 show similar behaviour except for the presence of a weak He I line at 1.0830 $\mu$m, NI lines at 1.2090 and 1.3450 $\mu$m, and 
the absence of the weak line of Mg at 1.1828 $\mu$m.  
The K band spectrum shows Br8 line, C I line at 1.9722 $\mu$m,  
He I line at 2.0581 $\mu$m and weakening of Na I line at 2.2056 $\mu$m compared to the March 13 data.

The next spectra taken on 2008 April 9 show He I, Pa$\gamma$ weak Fe II at 1.1126 $\mu$m, strong line of OI at 1.1287 $\mu$m, weak CI line 1.1748 $\mu$m, weak 
NI line at 1.2090, 1.2461, 1.3450 and CI line at 1.2569, 1.1653 $\mu$m, Pa$\beta$, OI line at 1.3164 $\mu$m. 
The K band spectrum shows Br8 line, C I line at 1.9722 $\mu$m, strengthening of He I line at 2.0581 $\mu$m, disappearance of Na I line at 2.2050 $\mu$m and
sudden appearance of the line at 2.0888 $\mu$m. In their study on 
nova V2615 Oph, Das et al. (2009) have predicted that this line can be an Fe II line excited by Ly$\alpha$ fluorescence. 

The $JHK$ band spectra obtained on 2008 April 12 are similar to the previous spectra but the Ca II infrared triplet, which was very strong in March, is nearly undetectable in this spectrum. The He I lines are now quite strong, and He II features are also starting to emerge.

\begin{figure*}
\begin{center}
 \includegraphics[width=5.5in, height=5.5in]{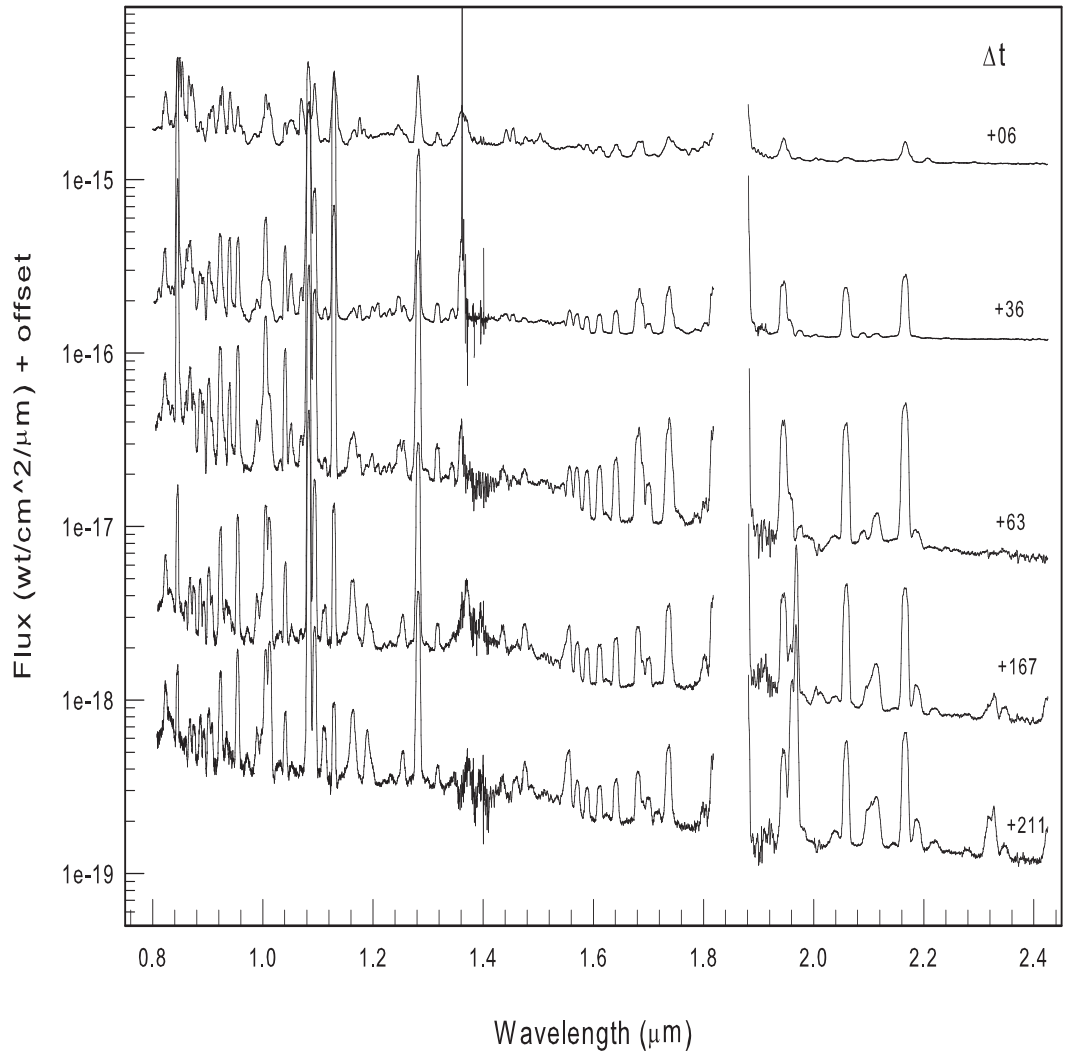}
  \caption{ The 0.8-2.4  $\mu$m portion of the infrared spectra of V2468 Cyg obtained from NASA Infrared Telescope Facility.  
  The time from outburst is given for each spectrum. These spectra are particularly important in showing the development of the coronal lines. 
  Note in particular the rise of the [Si VI] line at 1.9645 $\mu$m. At day 36, this feature is barely detectable on the red wing of the Br8 line. 
  But by day 167 it is the larger feature and by day 211 it is completely dominant. Other coronal lines that are present and display a similar development are 
  [S VIII] and [Ca VIII] at 0.9911 and 2.3205 $\mu$m, respectively.}
  %\label{ch5_4}
  \end{center}
  \end{figure*}

The spectrum taken on 2008 April 22 shows very strong He I line at 1.0830 $\mu$m, weak Pa$\gamma$ and absence of Fe II line,
strong emission of O I at 1.1287 $\mu$m, very weak NI lines at 1.2461, 1.2090, 1.3450 $\mu$m, Pa$\beta$, 
weak line of OI at 1.3164 $\mu$m. The H band spectrum shows most of the Brackett series lines, He I line at 1.7002 $\mu$m, Mg I line at 1.7109 $\mu$m. 
The C I line at 1.7808 $\mu$m is absent now.
The K band spectrum shows the strengthening of He I line at 2.0581 and the absence of C I line at 1.9722 $\mu$m, the Fe II line at 2.0888 $\mu$m and the appearance of He I line at 2.1120 $\mu$m.
The spectra taken on 2008 May 9 is very similar to the previous spectra. 
The spectra taken on 2008 May 15 are very similar to the previous spectra. 
But the disappearance of the Fe II line at 2.0888 $\mu$m is consistent with the rising ionization but is contrary to its observed persistence in the nova 
V2615 Oph (Das et al. 2009). 

The observations presented in Fig. 3 are of great importance as they show the development of the coronal lines. 
  We can see in particular the rise of the [Si VI] line at 1.9645 $\mu$m. This feature is barely detectable on the red wing of the Br8 line on 2008 April 12. 
  But by day 167 it is the larger feature and by day 211 it is completely dominant. The other coronal lines that are present and display a similar development are 
  [S VIII] and [Ca VIII] at 0.9911 and 2.3205 $\mu$m, respectively. 

The $LM$ band spectra taken on three epochs (see Fig. 4) show the emission lines of [Ca VIII], [Si VII] and H I lines. 
We do not find any evidence of thermal emission from dust. The absence of dust formation, together with the relatively rapid appearance of coronal lines support 
the notion that V2468 Cyg was an Fe IIb type nova.

\subsection{The reddening and distance of V2468 Cyg}

The relative strengths of the fluorescently excited lines of neutral oxygen have been used to determine the reddening in numerous novae and other emission line 
objects (Rudy et al. 1991). This makes use of the specific O I lines at 0.8446, 1.1287, and 1.3164 $\mu$m.  
For the first epoch  2008 March 13, there is a strong Ca II triplet that is partially blended with 0.8446 $\mu$m and the lines are broad with complex profiles.
This contributes to the uncertainty in the determination of reddening. By 2008 April 12, however, the OI lines are dominant and can be measured accurately. 
The reddening for the four epochs between 2008 April 12 and 2008 October 4 yield a value of E(B-V) = 0.77 $\pm$ 0.15 as reported in Rudy et al. (2008).
% This all contributes to the uncertainty in the reddening which is, nevertheless, significant. 
% By the 2008 April 12, however, the O I lines dominate their spectral region and can be measured accurately. 
% The reddening for the four epochs between 2008 April 12 and 2008 October 4 yield a best value of E(B-V) = 0.76 $\pm$ 0.15.
% 
This value is similar to the values
derived by Schwarz et al. (2009) and Tarasova (2013) $\sim$ 0.80 from the Balmer decrement, Chochol et al. (2012) $\sim$ 0.79 using B and V band light curve and 
Iijima and Naito (2011) $\sim$ 0.80 using the column density of hydrogen atoms. Using MMRD relation (della Valle \& Livio, 1995), we estimated the absolute 
magnitude at maximum M$_V$ and distance d to the nova $\sim$ -8.8 and $\sim$ 5.6 kpc, respectively where we used $t_2$ = 7.8 d from Iijima and Naito (2011) and 
our estimate of E(B-V) above. We also estimated the mass of the white dwarf $M_{WD}$ $\sim$ 1.1M$_\odot$ (Livio 1992). This suggests that nova V2468 Cyg may 
have been originated from a massive WD.

\subsection{A possible case of hybrid nova}

Williams(1992) has suggested a new class of novae which display characteristics of both ‘FeII’ and ‘He/N’ classes and these are referred to as ‘hybrid’ or ‘FeIIb” 
novae. The observed spectral lines of ‘hybrid’ novae are broader than the ones typically seen in ‘FeII’ nova and subsequently exhibit characteristics of ‘He/N’ 
novae like strong He/N lines. The observed FWHM of HI lines in V2468 Cyg in the early decline phase range from 1800 to 2300 km s$^{-1}$ (Fig. 6) which are 
larger compared
to those seen in typical ‘FeII’ novae. A remarkable and rapid increase in the strength of HeI lines starting from day 33 after the outburst (2008 April 9) is seen
as observed in He/N novae e.g. KT Eri (Raj et al. 2013). An extension of the optical classification scheme to the near-infrared by Banerjee and Ashok (2012) 
distinguishes these two classes by the presence of strong CI lines in “FeII’ novae and their absence in ‘He/N’ novae. The spectra of V2468 Cyg displayed in 
Figs. 2-4 clearly show that the strong CI lines seen in the initial spectra quickly fade away. A similar behaviour and spectral evolution was seen in V574 Pup 
that was classified  by Naik et al.(2010) as ‘hybrid’ nova. We thus suggest a ‘hybrid’ classification for V2468 Cyg. 
At the optical wavelengths the signatures of the ‘hybrid’ class of  novae are the spectral features of HeII 3923, 6683, 6891 \AA, NII 5001, 5479, 6482 \AA, and 
NI 7452\AA.  
The optical spectra of V2468 Cyg show the presence of NII 5679 \AA, line on 2008 March 10 (Iijima \& Naito 2011), traces of NII 5938 \AA, on 2008 March 23 
and NII 6482 \AA, on 2008 May 22 (Tarasova 2013). The presence of spectral lines of He and N at such early epochs is consistent with our suggestion that V2468 Cyg
belongs to the ‘hybrid’ class of nova.

An example of well studied nova that showed transition between FeII and He/N spectral  classes is the recent 2011 outburst of recurrent nova T Pyx. Unlike hybrid 
novae T Pyx showed transition from He/N to FeII spectral class (Izzo et al. 2012 and Surina et al. 2014). The high resolution spectroscopic 
observations of T Pyx by Shore et al. (2011) reported absorption line systems that show an accelerated displacement in velocity and they attribute these features 
to a fast moving tenuous outer envelope. The optical spectroscopic observations of V2468 Cyg by Chochol et al. (2012) show absorption components in the H$\alpha$ 
emission feature indicating the presence of an expanding outer envelope.

As Williams (2012) has pointed out, the FeII spectral features are formed in a post-outburst wind while the He/N spectral features arise in discrete expanding 
shells. It is likely that in case of hybrid novae the spectrum is dominated by the FeII lines with weak emission from He \& N in the initial phase and then changes 
over to the He/N class as the nova evolves to the nebular phase. The slow moving optically thick discrete shells with lower ionization are responsible for 
the spectral feature of FeII class in the initial phase and these are replaced by the optically thin shells with increased ionization, responsible for the 
spectral features of He/N class in later phase. The geometry of the ejecta could be another likely reason for the hybrid nature of selected novae. If the 
underlying white dwarf in the nova happens to be massive - a value of 1.1 M$_\odot$ is estimated for V2468 Cyg in section 3.3 and Tarasova (2013) -  
it may have magnetic field that 
could be strong enough to influence the expanding ejecta and change its geometry. The detection of any deviations from the spherical geometry by high angular
resolution observations of hybrid novae will give credence to the idea that their ejecta are concentrated in jets. 

The recent addition to the list of hybrid novae, namely, V5558 Sgr (Das et al. 2015) and V5588 Sgr (Munari et al. 2015) suggests that the transition between 
FeII and He/N class in novae may be more frequent than was believed in earlier years. 
% Williams (1992) has suggested a new class of novae which display characteristics of both classes
% and these are referred to as 'hybrid' or 'Fe IIb' novae. The typical FWHM of the H I lines in V2468 Cyg range from 1800 to 2300 km s$^{-1}$ (Fig. 6).
% This is large as compared to those seen in typical 'Fe II' novae. 
% This is illustrated in Figs 2-4 wherein the CI lines, characteristic of the Fe II class, quickly disappear and
% a remarkable and rapid increase in the strength of the He I lines is seen as expected in the He/N class of novae e.g. KT Eri (Raj et al. 2013). 
% A similar behaviour and spectral evolution is seen in the case of V574 Pup, another possible hybrid nova (Banerjee \& Ashok 2012, Naik et al. 2010).

\begin{figure*}
  \begin{center}
  \includegraphics[width=5.5in,height=5.5in]{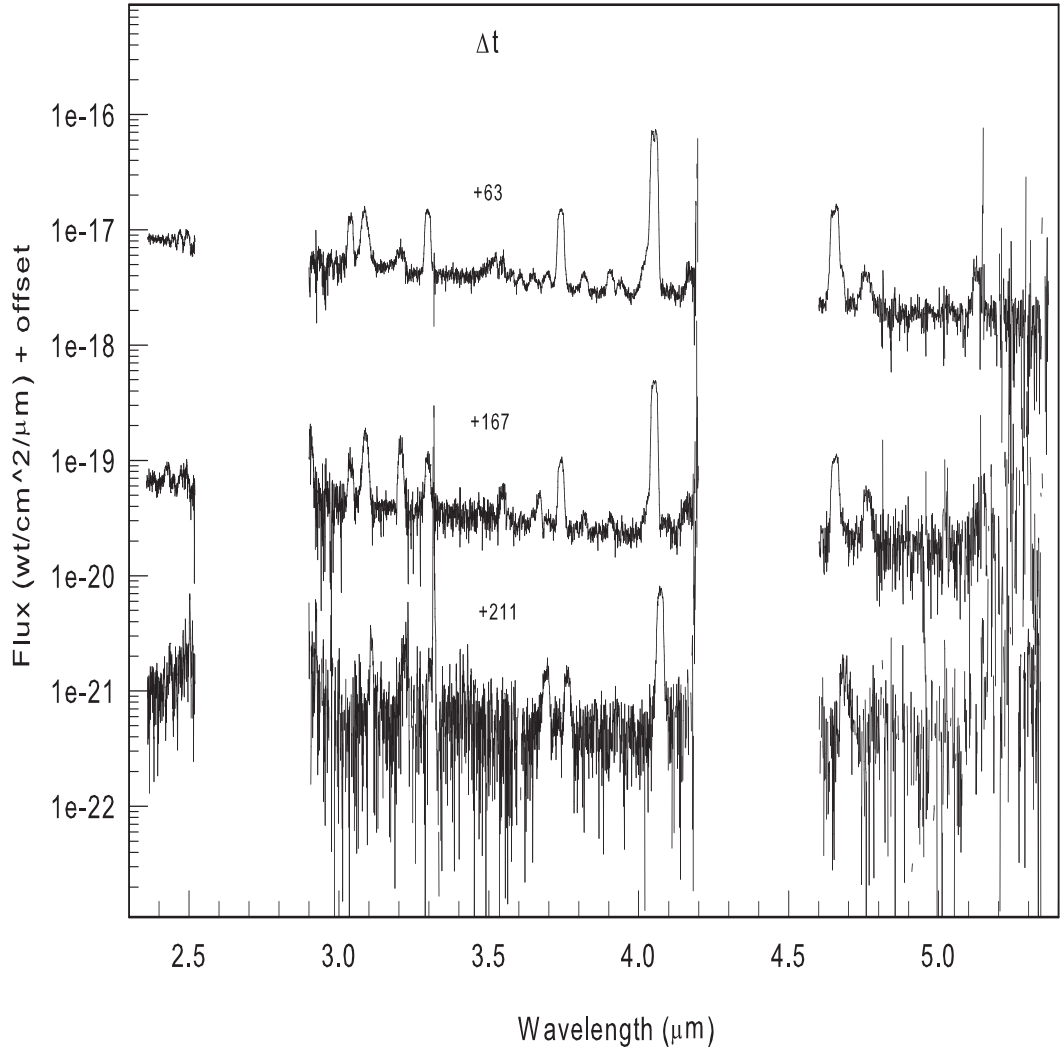}
 \caption{ The 2.3 to 5.2 ${\rm{\mu}}$m portions of the spectra V2468 Cyg taken from NASA Infrared Telescope Facility. 
 The time from outburst is given for each spectrum. All of the emission lines, with the exception of the coronal lines [Ca VIII] and [Si VII] that appear 
 in the day 167 spectrum, are H I features.  None of these spectra show any evidence of thermal emission from dust. The absence of dust formation, 
 together with the relatively rapid appearance of coronal lines support the notion that V2468 Cyg is an Fe IIb type nova.}
     %\label{ch5_5}
     \end{center}
  \end{figure*}

\begin{figure}
  \begin{center}
\includegraphics[width=3.0in,height=5.0in, clip]{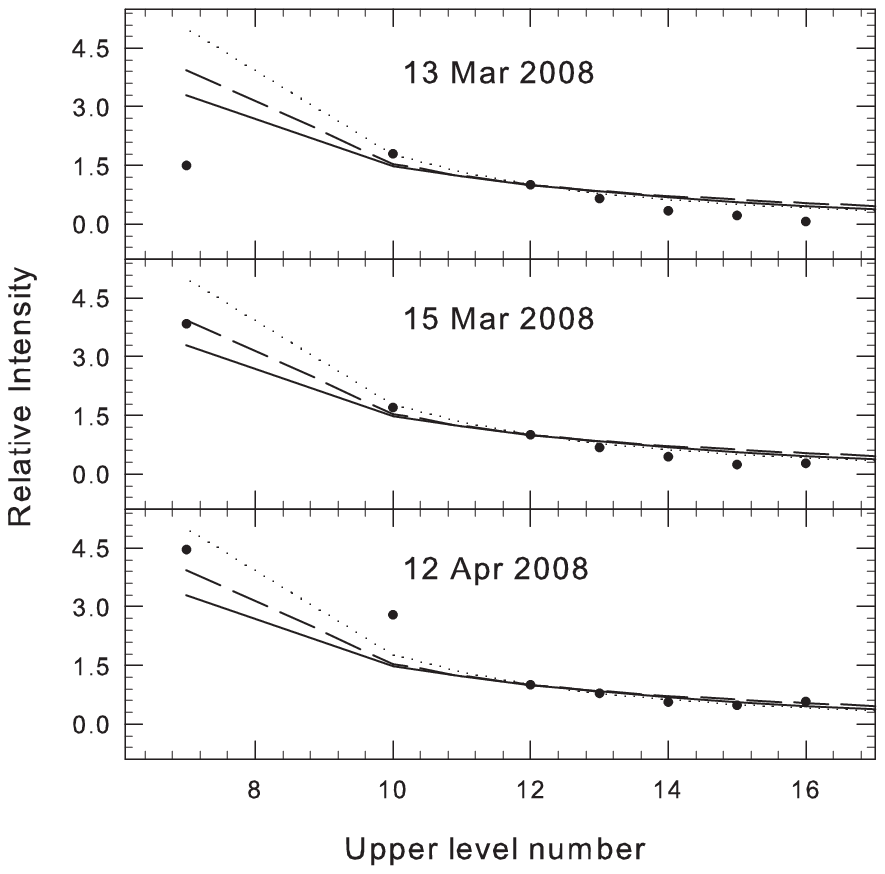}
  \caption{Recombination analysis for the hydrogen Brackett lines in V2468 Cyg on selected dates.
The abscissa is the upper level number of the Brackett series line transition. The line intensities are relative to that of Br 12. The Case B model
 predictions for the line strengths are also shown for a temperature of T = 10$^{4}$ K and electron densities of n$_e$ = 10$^{11}$ cm$^{-3}$
 (dotted line), 10$^{9}$ cm$^{-3}$ (solid line) and 10$^8$ cm$^{-3}$ (dashed line).}
  \label{ch5_6}
  \end{center}
  \end{figure}

\subsection{Recombination analysis of the H I lines and estimate of the ejecta mass}

We tried to estimate the ejecta mass by using recombination line analysis of H I lines and the representative results for three epochs of
our observations are shown in Fig. 5. However, 
we find that the strengths of these lines, relative to each other deviate considerably (specially for Br$\gamma$) from Case B values for 2008 March 13 and 
2008 April 12 clearly indicates that the observed line intensities are different from Case B values (Storey \& Hummer 1995). This is not surprising given the 
strength of the O I lines that are fluorescently excited by Ly$\beta$.  For this process to occur so productively, H$\alpha$ must be optically thick, which violates
the criterion for Case B conditions.  
We have plotted in Fig. 5 the observed relative strength of Brackett series lines with the line
strength of Br12 as unity along with the predicted values for three different recombination case B emissivity values from Storey \& Hummer (1995).
These predicted values cover a representative temperature of T = 10$^4$ K and the electron densities of n$_e$ = 10$^8$, 10$^9$ and 10$^{11}$ cm$^{-3}$.
High electron densities are considered because the ejecta material is dense in the early stages after the outburst as we do not see any auroral and nebular 
lines in this phase of observations.
Fig. 5 (middle panel) shows
that the observed line intensities clearly match from case B values in the initial phase i.e. on 2008 March 15 and started deviating afterwards.
 Specifically, Br${\rm{\gamma}}$, which becomes relatively stronger than the other Br lines showing thinning of the ejecta. 

For 2008 March 15, it is found that the observed data match well with the predicted recombination case B values of T = 10$^4$ K and
an electron density n$_e$ = 10$^{8}$ cm$^{-3}$. Following Banerjee et al. (2010), a constraint on the mass of the emitting gas can be obtained from:

\begin {equation}
M = (4 \pi d^2 (m_H)^2(fV/\epsilon))^{0.5}
\end {equation}
where $d$ is the distance derived in section 3.3, m$_H$ the proton mass, $f$ the observed flux in a particular line, $\epsilon$ the corresponding case B emissivity,
 V is the volume of the emitting gas. We calculate the volume of the nova shell following Mustel \& Boyarchuk (1970) for an estimate for its thickness and 
 clumpiness. 
%  which is [4/3$\pi$ $(vt)^3$ $\phi$], where $\phi$, $v$, and $t$ are the filling factor, velocity and time
% after outburst, respectively.
In the early stage of nova evolution the typical value of the filling factor vary from 10$^{-1}$ to 10$^{-2}$ (see Ederoclite et al. 2006 for more details). 
Cosidering a value of 10$^{-1}$ to 10$^{-2}$ for the filling factor we estimated the ejecta mass in the range 3 $\times$ 10$^{-6}$ - 10$^{-5}$M$_\odot$ which 
is similar to the values obtained by Ijima \& Naito (2011) and Tarasova (2013) from their optical observations. 
Although the ejecta mass is little closer to the mass of the ejecta seen in recurrent novae 
(e.g., Prialnik and Kovetz 1995) but V2468 Cyg can not be a recurrent nova (e.g., RS Oph, T Pyx) as it developed at a more 
leisurely rate and stayed bright for a very long time, displaying moderately strong emission lines more than four years after outburst. 
This  behavior argues for a more massive shell and persistent, low-level burning of material on the WD surface. 

\begin{figure}
\begin{center}
\includegraphics[width=3.0in,height=5.0in, clip]{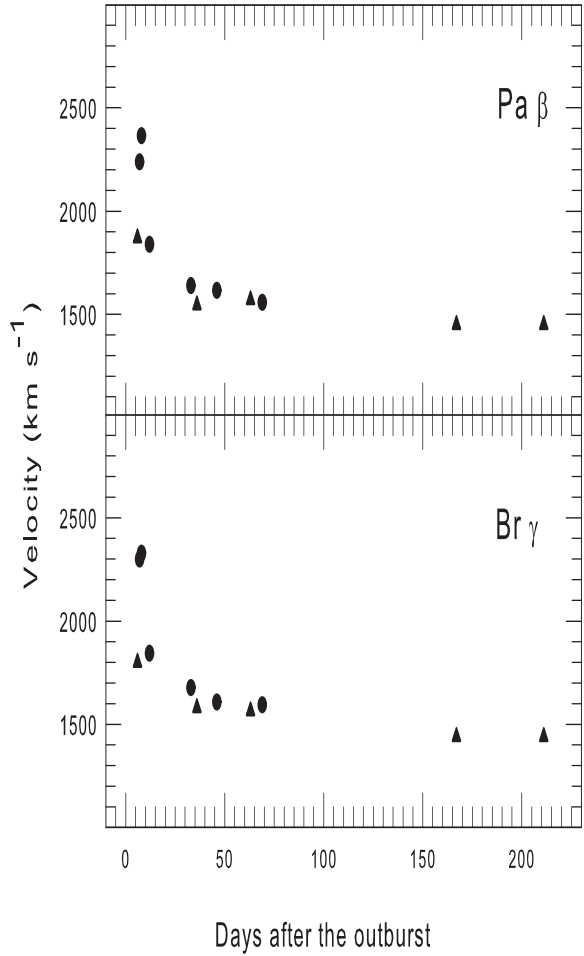}
  \caption{The FWHM for Pa$\beta$ and Br$\gamma$ lines for different epochs are shown. The solid circles and the triangles represent the data from Mt. Abu and 
  NASA Infrared Telescope Facility, respectively.}
  \label{ch5_9}
  \end{center}
  \end{figure}

\section{Summary}

We have presented the infrared spectroscopy and photometry of nova V2468 Cyg which erupted in early-March 2008. 
The infrared spectra indicate that the nova is a Fe IIb or hybrid class nova which shows Fe II type behaviour in early stages and then changes to He/N type 
later on. The reddening and distance to the nova is also calculated. Recombination analysis is used to estimate the mass
of the gaseous component of the ejecta in V2468 Cyg. 

\section{Acknowledgements}

The research work at Physical Research Laboratory is funded by the Department of Space, Government of India. The data from the Infrared Telescope Facility (IRTF) 
were acquired with the SpeX 0.8-5.5 $\mu$m spectrograph (Rayner et al. 2003)
and were reduced to absolute fluxes using the facility furnished Spextool software (Cushing et al. 2004, Vacca et al. 2003). 
The authors are thankful to the anonymous referee for the helpful comments that improved the manuscript.

\end{document}